\documentclass[reprint,superscriptaddress,nofootinbib,amsmath,amssymb,aps,pra]{revtex4-1}
\usepackage[latin9]{inputenc}
\usepackage{amsmath}
\usepackage{graphicx}
\usepackage[skip=10pt plus1pt, indent=10pt]{parskip}
\usepackage[unicode=true,
 bookmarks=false,
 breaklinks=false,pdfborder={0 0 1},backref=section,colorlinks=false]
 {hyperref}
\makeatletter

\DeclareTextSymbolDefault{\textquotedbl}{T1}

\usepackage{dcolumn}
\usepackage{physics}
\usepackage{braket}
\usepackage{bm}

\makeatother

\begin{document}
\global\long\def\abs#1{\left|#1\right|}%
\global\long\def\ket#1{\left|#1\right\rangle }%
\global\long\def\bra#1{\left\langle #1\right|}%
\global\long\def\half{\frac{1}{2}}%
\global\long\def\partder#1#2{\frac{\partial#1}{\partial#2}}%
\global\long\def\comm#1#2{\left[#1,#2\right]}%
\global\long\def\vp{\vec{p}}%
\global\long\def\vpp{\vec{p}'}%
\global\long\def\dt#1{\delta^{(3)}(#1)}%
\global\long\def\Tr#1{\textrm{Tr}\left\{  #1\right\}  }%
\global\long\def\Real#1{\mathrm{Re}\left\{  #1 \right\}  }%
\global\long\def\braket#1{\langle#1\rangle}%
\global\long\def\escp#1#2{\left\langle #1|#2\right\rangle }%
\global\long\def\elmma#1#2#3{\langle#1\mid#2\mid#3\rangle}%
\global\long\def\ketbra#1#2{|#1\rangle\langle#2|}%

\title{Bright and dark solitons in a photonic nonlinear quantum walk:\\ lessons from the continuum}
\author{Andreu Anglés-Castillo}
\email{andreu.angles-castillo@uv.es}

\affiliation{Departament de Física Teòrica \& IFIC, Universitat de València-CSIC,
46100 Burjassot (València) Spain}
\author{Armando Pérez}
\affiliation{Departament de Física Teòrica \& IFIC, Universitat de València-CSIC,
46100 Burjassot (València) Spain}
\author{Eugenio Roldán}
\affiliation{Departament d\textquoteright Òptica i d\textquoteright Optometria
i Ciències de la Visió, Universitat de València, Dr. Moliner 50, 46100-Burjassot}
\begin{abstract}
We propose a nonlinear quantum walk model inspired in a photonic implementation in which the polarization state of the light field plays the role of the coin-qubit. In particular, we take profit of the nonlinear polarization rotation occurring in optical media with Kerr nonlinearity, which allows to implement a nonlinear coin operator, one that depends on the state of the coin-qubit. We consider the space-time continuum limit of the evolution equation, which takes the form of a nonlinear Dirac equation. The analysis of this continuum limit allows us to gain some insight into the existence of different solitonic structures, such as bright and dark solitons. We illustrate several properties of these solitons with numerical calculations, including the effect on them of an additional phase simulating an external electric field.
\end{abstract}
\maketitle

\section{Introduction}

The quantum walk (QW) is a powerful toolbox with many applications.
It can be shown to constitute a universal model of computation \cite{Childs2009,CGW13,Lovett2010}
with algorithmic applications, such as search problems \cite{Childs2004,Tulsi2008,MNRS11,Ambainis2013,FG2014,RGAM20}
or element distinctness \cite{Amb07a}. QWs manifest into two main
categories. Continuous-time QWs (CQWs) are described by a local Hamiltonian
originated from the adjacency matrix on some graph with a time evolution
which is dictated by the Schrödinger equation, while discrete-time
QWs (DQWs) are defined by a unitary evolution operator which relates
two consecutive time instants in a stroboscopic way. Another important
difference is that, in the case of DQWs, the Hilbert space associated
to the graph needs to be enlarged with an additional degree of freedom
(the so called ``coin'' space). In spite of this different formulation,
it is possible to establish a connection between CQWs and DQWs \cite{Strauch06a,Strauch07a,ChildsCD2009,PP16,Schmitz2016}.
In this work, we will concentrate on DQWs.

From a physical point of view, DQWs have also been used for the simulation of various physical theories and phenomena. Many of these applications are motivated from the fact that, under the appropriate conditions, the continuum limit of DQWs is the Dirac equation. In this way, DQWs can be used to simulate spin-1/2 particles in both external Abelian \cite{DDMEF12a,AD16a,AD16b} and non-Abelian \cite{AMBD16} gauge fields. Such simulations can also be applied to relativistic gravitational fields \cite{DMD13b,DMD14,AD17,Arrighi_curved_1D_15,AF17}. DQWs also show additional interesting phenomena \cite{Molfetta2016,BlochOscillationQW,MrquezMartn2017}. 
In addition, QWs have been implemented using different setups \cite{book_Manouchehri}, such as photons in various optical devices \cite{Trompeter06,Schreiber10a,Peruzzo10a,Kitagawa2012,Sansoni11a,sciarrino12,BNFO2015_synthetic_gauge_fields}, atoms trapped in arrays of light \cite{GASW13}, ion traps \cite{Bruzewicz2019}, or superconducting qubits \cite{Kjaergaard2020}.

In this work, we analyze a variant of the DQW which introduces nonlinearities on the angle of the coin operator, and shows some similar phenomena as in the Non-Linear Optical Galton Board (NLOGB) model introduced in \cite{Navarrete07}, where such nonlinearities appeared as phases on the different components of the dynamical map. The main result in \cite{Navarrete07} was the appearance of soliton-like structures with a rich phenomenology that can be controlled by varying the coupling strength to the nonlinear Kerr medium. In the model we propose below, we observe the formation of bright solitons, as in the NLOGB, and also of dark solitons. We are able to connect these solutions with the continuum space-time limit of the QW, which can be easily obtained. We analyze numerically some aspects of the dynamics of these solitonic structures, including the effect of an additional electric field, and we also show that these solitons do not appear in the two-dimensional case.

This paper is organized as follows. In Sect. \ref{sec:Overview-on-linear} we first recall the setup for the linear DQW, and review the different proposals to account for a nonlinear DQW. In Sect. \ref{sec:Model} we introduce our own proposal, and we discuss its experimental implementation based on nonlinear Kerr optical media. Sect. \ref{sec:Continuum-limit} is devoted to the analysis of the continuum space-time limit, which is afterwards illustrated by our numerical calculations in Sect. \ref{sec:Soliton-formation}. We conclude in Sect. \ref{sec:Conclusions} by summarizing our main
findings.

\section{\label{sec:Overview-on-linear}Overview on linear and nonlinear DQWs}

We start by briefly revisiting the standard (linear) and nonlinear
models to describe the DQW for a walker on a one-dimensional lattice.

\subsection{Linear DQW}

Let us consider a particle (the walker) which can move along a discrete
lattice with positions $x=j\epsilon,\,\,j\in\mathbb{Z}$, with $\epsilon$
the lattice spacing. A position Hilbert space $\mathcal{H}_{x}$ is
associated to this system, which is spanned by the basis $\{\ket x\}$,
with $x$ the lattice positions. As mentioned in the Introduction, we also need an additional degree of freedom that defines the coin
Hilbert space $\mathcal{H}_{c}$, and will be spanned by two orthogonal
states $\{\ket{\uparrow},\ket{\downarrow}\}$. The total Hilbert space
is, therefore, the tensor product $\mathcal{H}=\mathcal{H}_{x}\otimes\mathcal{H}_{c}$,
and the basis that spans the whole space is $\{\ket x\otimes\ket{\uparrow\downarrow}\}_{j\in\mathbb{Z}}$.
For reasons that will be explained in Sect. \ref{sec:Continuum-limit},
we define a time step evolution of the walker using the same amount
$\epsilon$, i.e. the state $\left\vert \psi_{t}\right\rangle $ at
a given time $t$ evolves as
\begin{equation}
\left\vert \psi_{t+\epsilon}\right\rangle =U\left\vert \psi_{t}\right\rangle~,
\label{eq:EvolutionUnitary}
\end{equation}
with $U$ the evolution operator. The operator $U$ is the composition
of two unitary operators,
\begin{equation}
U=SC~,
\end{equation}
where $C=\mathbb{I}\otimes R$ is the coin operator acting on $\mathcal{H}_{c}$.
In the latter equation, $S$ represents the conditional displacement
operator, which can be formally written as
\begin{equation}
   S=e^{-i\sigma_{z}\hat{p}}~,
\end{equation}
with $\hat{p}$ the lattice quasi-momentum operator, and $\sigma_{i},\,\,i=x,y,z$
the Pauli matrices acting on the $\{\ket{\uparrow},\ket{\downarrow}\}$
states. As for the operator $R$, it will be represented by a $2\times2$
unitary matrix. An example is given by
\begin{equation}
R\equiv e^{-i\theta\sigma_{y}}=\left(\begin{array}{cc}
\cos\theta & -\sin\theta\\
\sin\theta & \cos\theta
\end{array}\right)~.
\end{equation}

In what follows, we will set $\epsilon=1$, so that $x=j$. However,
we will need to restore this parameter in Sect. \ref{sec:Continuum-limit},
in order to derive the continuum limit of Eq. (\ref{eq:EvolutionUnitary}).

In terms of the tensor basis in $\mathcal{H}$, one can expand $\left\vert \psi_{t}\right\rangle $
as follows:
\begin{equation}
\left\vert \psi_{t}\right\rangle =\sum_{x}\left[u_{t,x}\left\vert x,\uparrow\right\rangle +d_{t,x}\left\vert x,\downarrow\right\rangle \right]~.
\end{equation}
In other words, the corresponding spinor is
\begin{equation}
\Psi(t,x)\equiv\Braket{x|\psi_{t}}=u_{t,x}\ket{\uparrow}+d_{t,x}\ket{\downarrow}\equiv\begin{pmatrix}u_{t,x}\\
d_{t,x}
\end{pmatrix}~.
\end{equation}
Finally, the operator $S$ takes the form
\begin{equation}
S=\sum_{x}\big(\ket{x+1}\bra x\otimes\ket{\uparrow}\bra{\uparrow}+\ket{x-1}\bra x\otimes\ket{\downarrow}\bra{\downarrow}\big)~.
\end{equation}

\subsection{Nonlinear DQW}

The discrete nonlinear QW was not first introduced as such, but as
a nonlinear Optical Galton Board \cite{Navarrete07}, mainly because
\textquotedbl nonlinear quantum walk\textquotedbl{} is close to be
an oxymoron, being quantum mechanics a linear theory; however, the
term Non-linear Quantum Walk (NLQW) has made its way through the literature
and we adhere to it, but we must keep in mind that the waves used
in a NLQW cannot be true quantum wave-functions but some other type
of waves.

The NLOGB is a coined DQW on the line in which the wavefunction acquires
an additional coin-state-dependent nonlinear phase $\phi_{c,NL}$
depending on the probability as $\phi_{c,NL}=i2\pi\alpha\left\vert c_{t,x}\right\vert ^{2}$
with $c=u,d$ and $\alpha$ the nonlinearity strength. This is equivalent
to either (i) replacing the standard QW coin operator $R$ by the
inhomogeneous nonlinear coin operator
\begin{equation}
R_{t,x}=\left(\begin{array}{cc}
e^{i2\pi\alpha\left\vert u_{t,x}\right\vert ^{2}}\cos\theta & -e^{i2\pi\alpha\left\vert d_{t,x}\right\vert ^{2}}\sin\theta\\
e^{i2\pi\alpha\left\vert u_{t,x}\right\vert ^{2}}\sin\theta & e^{i2\pi\alpha\left\vert d_{t,x}\right\vert ^{2}}\cos\theta
\end{array}\right)~,
\end{equation}
or to (ii) generalizing the conditional displacement operator as 
\begin{eqnarray}
	S  =  \sum_{x=-L}^{L}& e^{i2\pi\alpha\left\vert u_{t,x}\right\vert ^{2}}\ket{x+1}\bra x\otimes\ket{\uparrow}\bra{\uparrow} \nonumber \\
   & +e^{i2\pi\alpha\left\vert d_{t,x}\right\vert ^{2}}\ket{x-1}\bra x\otimes\ket{\downarrow}\bra{\downarrow}~.
\end{eqnarray}
In \cite{Navarrete07}, the NLOGB was numerically studied and the
existence of solitons and of rich spatio-temporal dynamics, including
chaotic behaviour, was shown. The NLOGB was later experimentally implemented
by Wimmer et al. \cite{Wimmer13} in a system involving the propagation
of light pulses in optical fibres, an implementation in which the
displacement operation consists in delaying or advancing the pulses,
so that the QW occurs along the physical time dimension. More recently,
the same group made a proposal of NLQW in optical mesh lattices \cite{Wimmer21},
see also the related paper \cite{Price22}, a system that has been
recently revisited by Yue et al. \cite{Yue23}. The NLOGB model has
also been the subject of several theoretical studies, including the
study of its continuous limit as a nonlinear Dirac equation \cite{DiMolfetta15,Lee15,Maeda19b}.
Further numerical studies by Buarque and coworkers centred on self-trapping
\cite{Buarque20}, breathing dynamics \cite{Buarque21}, and rogue
waves \cite{Buarque22}. There has also been made a rigorous mathematical
study of the discrete model \cite{Maeda18,Maeda19} including the
demonstration of long term soliton stability \cite{Maeda22}. Recently,
the NLOGB has been extended to three-state coins \cite{Falcao22},
and generalized to include the effect of perturbing potential barriers
\cite{Passos22}.

Moreover, NLQWs different from the NLOGB have been proposed. Shikano
et al. \cite{Shikano14} proposed a NLQW in which the nonlinearity
is due to a feed-forward quantum-coin mechanism such that the coin
elements become $\cos\theta_{t,x}=\left\vert u_{t-1,x+1}\right\vert +i\left\vert d_{t-1,x-1}\right\vert $
and $\sin\theta_{t,x}=\sqrt{1-\left\vert \cos\theta_{t,x}\right\vert ^{2}}$.
Lee et al. show how the dynamics of a nonlinear Dirac particle can
be simulated by NLQWs with a measurement-based feed-forward scheme,
slightly different from that of Shikano et al., considering both Gross-Neveu
and Thirring types of nonlinear couplings. Gerasimenko et al. \cite{Gerasimenko16}
introduced the nonlinearity through the operator $\exp\left[-i\kappa\left(\left\vert u_{t,x}\right\vert ^{2}-\left\vert d_{t,x}\right\vert ^{2}\right)\sigma_{y}\right]$,
so that the nonlinear phase-shift depends on the "population difference"
$\left\vert u_{t,x}\right\vert ^{2}-\left\vert d_{t,x}\right\vert ^{2}$,
and concentrated in the study of the influence of zero modes on the formation
of solitonic structures in the continuum limit. In \cite{Mochizuki20}
the work in \cite{Gerasimenko16} was generalized by using mathematical
techniques appropriate to Floquet systems, which allowed for the
finding of new bifurcations.

Another alternative is that of Mendonça et al. \cite{Mendonca20},
who propose a nonlinear displacement operator of the form
\begin{equation}
\begin{split}
S & =\sum_{x=-L}^{L}\frac{1}{\sqrt{1+\alpha P_{x}}}\left[\left\vert x+1,\uparrow\right\rangle \left\langle x,\uparrow\right\vert +\alpha P_{x}\left\vert x,\uparrow\right\rangle \left\langle x,\uparrow\right\vert \right]\\
 & +\sum_{x=-L}^{L}\frac{1}{\sqrt{1+\alpha P_{x}}}\left[\left\vert x-1,\downarrow\right\rangle \left\langle x,\downarrow\right\vert +\alpha P_{x}\left\vert x,\downarrow\right\rangle \left\langle x,\downarrow\right\vert \right]~,
\end{split}
\end{equation}
with $P_{x}=\left\vert u_{t,x}\right\vert ^{2}+\left\vert d_{t,x}\right\vert ^{2}$.
They numerically find and describe a variety of nonlinear phenomena, which were further studied in \cite{Gong22}.

As for Mallick et al. \cite{Mallick22}, they use the nonlinear map
\begin{align}
	u_{t+1,x}&=\cos\theta u_{t,x-1}+e^{i\phi_{x-1}(t)}\sin\theta d_{t,x-1}~, \\
	d_{t+1,x}&=-e^{-i\phi_{x+1}(t)}\sin\theta u_{t,x+1}+\cos\theta d_{t,x+1}~,
\end{align}
with $\phi_{x}=\gamma P_{x}+\eta_{x}$, where $\eta_{x}$ is a noise
term, and study the breakdown of Anderson localization
induced by the nonlinearity. Finally, in \cite{Zeng22} single atoms
are proposed as nonlinear beam-splitters in their proposal of a NLQW.

Closely related studies are those by Solntsev et al. \cite{Solntsev14},
who incorporate biphoton generation --an intrinsic nonlinear process--
in a photonic wave-guide array and study the potential of the system
to generate entangled light, but their quantum walk is linear; Verga
\cite{Verga17} studies edge-states in a QW with both linear and nonlinear
disorder; Bisio et al. \cite{Bisio18} analytically diagonalize a
discrete-time on-site interacting fermionic cellular automaton in
the two-particle sector; Adami et al. \cite{Adami19} study a NLQW
naturally induced by a quantum graph with nonlinear delta potentials;
Templeman et al. \cite{Tempelman21} study topological protection
in a strongly nonlinear interface lattice;\ and Held et al. \cite{Held22}
introduce Gaussian QWs, which are NLQWs in which the coins are substituted
by two-mode squeezers. As stressed by the authors, this kind of NLQWs
directly lead to accessible quantum phenomena, rendering possible
the quantum simulation of nonlinear processes.

We must also mention works on continuous time NLQWs. In \cite{Piskovski08}
the destruction of Anderson localization by nonlinearity is studied
through discrete Anderson nonlinear Schrödinger equations that correctly
describe the one-dimensional disordered waveguide lattices used in
the experiments of Lahini et al. \cite{Lahini08}. But most studies
are related to the problem of database searching. Ebrahimi Kahou et
al. study this problem with coupled discrete nonlinear Schrödinger
equations, and discuss the implementability of the model with BECs
\cite{Kahou13}. Meyer and Gong study quantum search with the Gross-Pitaevskii
equation \cite{Meyer13,Meyer14} concluding that it solves the unstructured
search problem more efficiently than does the Schrödinger equation,
because it includes a cubic nonlinearity, and Chiew et al. \cite{Chiew19}
demonstrate that the nonlinear quantum search can be more efficient
than quantum search for graph comparison. Di Molfetta et al. \cite{DiMolfetta20}
generalize the Meyer-Gross algorithm to two dimensions finding a clear
advantage over the linear QW. Finally, in \cite{Becerra23} the thresholds
between modulational stability, rogue waves and soliton regimes are
studied with coupled nonlinear Schrödinger equations with on site
saturating nonlinearity.

In the present paper, we introduce an alternative formulation of the
NLQW appropriate for light polarization qubits propagating in Kerr
media. Specifically, we introduce a nonlinear coin in which the rotation
angle is given by $\theta=\theta_{0}+\theta_{NL}$ with $\theta_{0}$
constant and $\theta_{NL}$ depending on the light polarization state,
hence in the coin state. 

\section{\label{sec:Model}Model}

\subsection{NLQW coin and map}

The nonlinear Quantum Walk (QW) we propose introduces the
non-linearity in the coin operator. The unitary operator $R$ is now
defined in a way that depends on the state of the walker 
\begin{equation}
R(\theta_{t,x})=e^{-i\theta_{t,x}\sigma_{y}}=\begin{pmatrix}\cos\theta_{t,x} & -\sin\theta_{t,x}\\
\sin\theta_{t,x} & \cos\theta_{t,x}
\end{pmatrix}~,
\end{equation}
where the angle of rotation is given by 
\begin{equation}
\theta_{t,x}=\theta_{0}+\alpha|u_{t,x}||d_{t,x}|\sin\delta_{t,x}~,\label{eq:NLangle}
\end{equation}
where we explicitly expressed the upper and lower components given
by their modules and complex angle as $u_{t,x}=|u_{t,x}|e^{i\varphi_{t,x}^{u}}$,
$d_{t,x}=|d_{t,x}|e^{i\varphi_{t,x}^{d}}$ and $\delta_{t,x}=\varphi_{t,x}^{u}-\varphi_{t,x}^{d}$
is the phase difference. If we write how the components evolve explicitly
after each time step, we get 
\begin{equation}\label{eq:MapQWNL}
\begin{split}u_{t+1,x} & =\cos(\theta_{t,x-1})u_{t,x-1}-\sin(\theta_{t,x-1})d_{t,x-1}~,\\
d_{t+1,x} & =\sin(\theta_{t,x+1})u_{t,x+1}+\cos(\theta_{t,x+1})d_{t,x+1}~.
\end{split}
\end{equation}

\subsection{Experimental proposal}

In proposing the nonlinear rotation term in Eq. (\ref{eq:NLangle}),
we are thinking in a QW photonic platform, using light-polarization
qubits \cite{PhysRevA.68.020301,Schreiber_2012}, and including optical
media with Kerr-type nonlinearity. It is well known that in an isotropic
Kerr medium the normal modes of propagation are circularly polarized, and their corresponding indexes of refraction are given by \cite{Boyd}
\begin{equation}
n_{\pm}=n_{0}+\frac{1}{2n_{0}}\left[A\left\vert E_{\pm}\right\vert ^{2}+\left(A+B\right)\left\vert E_{\mp}\right\vert ^{2}\right]~,
\end{equation}
where the subscripts $+$ and $-$ make reference to right- and left-circular
polarization, respectively, $n_{0}$ is the linear refractive index,
and $A$ and $B$ are the Maker-Terhune coefficients for the nonlinear
medium, whose ratio depends on the specific physical mechanism responsible
for the Kerr effect (e.g., $A=B$ for nonlinear electronic response).
In such a medium, the phenomenon of nonlinear polarization rotation
occurs, by means of which a polarized monochromatic-wave that propagates
a distance $z$ along the nonlinear medium has the expression
\begin{equation}
\begin{split}
	\vec{E}\left(z\right)=&E_{+}\hat{\sigma}_{+}+E_{-}\hat{\sigma}_{-}\\
	=&\left(A_{+}\hat{\sigma}_{+}e^{i\theta_{NL}}+A_{-}\hat{\sigma}_{-}e^{-i\theta_{NL}}\right)e^{ik_{m}z}~,
\end{split}
\end{equation}
where $k_{m}=\left(n_{+}+n_{-}\right)\omega/2c$ is the mean propagation
constant, $\hat{\sigma}_{\pm}=\left(\hat{x}\pm i\hat{y}\right)/\sqrt{2}$
are the unit circular polarization vectors, and 
\begin{equation}
\theta_{NL}=\frac{1}{2}\left(n_{+}-n_{-}\right)\frac{\omega}{c}z=\frac{B}{4n_{0}}\left(\left\vert E_{-}\right\vert ^{2}-\left\vert E_{+}\right\vert ^{2}\right)\frac{\omega}{c}z~.
\end{equation}
This expression shows that the polarization state of the light undergoes a rotation $\theta_{NL}$ after the propagation, the circular components changing from $\hat{\sigma}_{\pm}$ to $\hat{\sigma}_{\pm}^{\prime} = \hat{\sigma}_{\pm}e^{\pm i\theta_{NL}}$, and the linear components of the polarization passing from $\left(x,y\right)$ at the entrance to
\begin{align}
x^{\prime} & =x\cos\theta_{NL}-y\sin\theta_{NL}~,\\
y^{\prime} & =x\sin\theta_{NL}+y\cos\theta_{NL}~,
\end{align}
at the exit. Notice that for linearly polarized light $\left\vert E_{-}\right\vert ^{2}=\left\vert E_{+}\right\vert ^{2}$,
and hence $\theta_{NL}=0$. Notice also
that the rotation does not change the proportion $\left\vert E_{-}\right\vert /\left\vert E_{+}\right\vert $.

Now, we take the linear polarization components as the coin state
basis, so that the displacement operator acts on these linear components.
It is then necessary the use of the additional standard coin rotation
$\theta_{0}$ in Eq. (\ref{eq:NLangle}), because, after the action of the displacement
operator, the field is linearly polarized at the displaced positions,
and we have seen that there is no nonlinear rotation for linear polarizations,
which means that the nonlinear coin would not act but in the first
step. Finally, by writing $\theta_{NL}$ in terms of the linear polarization
components of the field, one arrives at the expression in Eq. (\ref{eq:NLangle}).

\section{Continuum limit\label{sec:Continuum-limit}}

The continuum limit of the QW is obtained by retaining the lowest order,
i.e. $\mathcal{O}(\epsilon)$, in the unitary evolution
defined by Eq.~\eqref{eq:EvolutionUnitary}. To this purpose, we
need to restore the parameter $\epsilon$ both in the lattice spacing
and in the time step. The continuum limit of this quantum walk can
be obtained following the standard method. Details and definitions
are given in Appendix \ref{app:ContLimit}.

The non-linear Dirac equation obtained from this limit reads 
\begin{equation}
\left[i\gamma^{\mu}\partial_{\mu}-m(\Psi(t,x))\right]\Psi(t,x)=0~,\label{eq:DiracNL}
\end{equation}
where the Dirac matrices are $\gamma^{0}=\sigma_{y}$ and $\gamma^{1}=i\sigma_{x}$
and the mass term is given by 
\begin{equation}
m(\Psi(t,x))=\tilde{\theta}_{0}-\frac{\tilde{\alpha}}{2}\Psi^{\dagger}(t,x)\sigma_{y}\Psi(t,x)~,
\end{equation}
where we defined the rescaled angle $ \epsilon \tilde \theta_0 = \theta_0$ and nonlinearity parameter $\epsilon \tilde \alpha = \alpha$.
We notice that this mass term is different from that obtained from
the NLOGB in \cite{DiMolfetta15}. We can write this equation in terms
of the spinor components $\Psi\left(t,x\right)=\left(u\left(t,x\right),d\left(t,x\right)\right)^{T}$
as
\begin{align}
	\left(\partial_{x}+\partial_{t}\right)u+\tilde{\theta}d & =0~,\\
	\left(\partial_{x}-\partial_{t}\right)d+\tilde{\theta}u & =0~,
\end{align}
with
\begin{equation}
\tilde{\theta}\equiv\tilde{\theta}_{0}+i\frac{\tilde{\alpha}}{2}\left(u^{\ast}d-ud^{\ast}\right)~,
\end{equation}
and we alleviated the notation by not writing the explicit spatial
and time dependence of each component. This system of equations can
be rewritten in terms of the modulus and phases of the spinor components
$u=|u|e^{i\varphi_{u}}$ and $d=|d|e^{i\varphi_{d}}$. After defining
\begin{equation}
	\delta=\varphi_{u}-\varphi_{d}~,\ \ \ \ \sigma=\varphi_{u}+\varphi_{d}~,
\end{equation}
one easily gets
\begin{equation}\label{ecs1}
\begin{split}
\partial_{x}\sigma+\partial_{t}\delta & =-\tilde{\theta}\frac{|u|^{2}-|d|^{2}}{|u||d|}\sin\delta~,\\
\partial_{x}\delta+\partial_{t}\sigma & =\tilde{\theta}\frac{|u|^{2}+|d|^{2}}{|u||d|}\sin\delta~,\\
\left(\partial_{x}+\partial_{t}\right)|u| & =-\tilde{\theta}|d|\cos\delta~,\\
	\left(\partial_{x}-\partial_{t}\right)|d| & =-\tilde{\theta}|u|\cos\delta~,
\end{split}
\end{equation}
with 
\begin{equation}
	\tilde{\theta}\equiv\tilde{\theta}_{0}+\tilde{\alpha}|u||d|\sin\delta~.
\end{equation}

\subsection{Homogeneous stationary solutions}
\label{sec:stability}

In order to gain some insight into the solutions of the system, we
first study the homogeneous stationary solutions and their stability. Let us focus on the last two equations of Eq. (\ref{ecs1}), which can be related as
\begin{equation}
\partial_{t}(|u|^{2}+|d|^{2})=-\partial_{x}(|u|^{2}-|d|^{2})~,
\end{equation}
which for stationary solutions
\footnote{A stationary solution refers to a static probability distribution
where the phases of the spinor components can still have a time dependence.}, and from the condition of normalization of the wavefunction, implies that $|u|=|d|$, and hence $\partial_{t}|u|=\partial_t|d|=0$, which implies in the last two equations of \eqref{ecs1} that $\delta$ is also time independent.

We can now define the intensity $I=|u|^{2}=|d|^{2}$
to rewrite Eq.~\eqref{ecs1} as 
\begin{align}
	\partial_{x}\sigma& =0~,\label{eqn:estat1}\\
	\partial_{x}\delta+\partial_{t}\sigma & =2\tilde{\theta}\sin\delta~,\label{eqn:estat2}\\
	\partial_{x}I & =-2I\tilde{\theta}\cos\delta~,\label{eqn:estat3}
\end{align}
with $\tilde{\theta}=\tilde{\theta}_{0}+\tilde{\alpha}I\sin\delta$. To further derive homogeneous stationary solutions, we impose $\partial_{x}I=\partial_{x}\delta=0$, which implies that $\partial_{t}\sigma=2\tilde{\theta}\sin\delta$
and $2I\tilde{\theta}\cos\delta=0$. The latter equation admits several
solutions: (i), $I=0$ (trivial solution); (ii) $\tilde{\theta}=0$,
which implies $I\sin\delta=-\tilde{\theta}_{0}/\tilde{\alpha}$ that
only exists for $I>\left\vert \tilde{\theta}_{0}/\tilde{\alpha}\right\vert $
and for which $\partial_{t}\sigma=0$; and (iii), the two solutions
$\delta=\pm\pi/2$ that exist for any value of $I$ and for which
$\partial_{t}\sigma=\pm2\tilde{\theta}=2\left(\tilde{\alpha}I\pm\tilde{\theta}_{0}\right)$.
Notice that for positive (negative)$\ \tilde{\theta}_{0}/\tilde{\alpha}$,
solution $\delta=-\pi/2$ ($\delta=\pi/2$) merges with solution $\tilde{\theta}=0$
for $I=\left\vert \tilde{\theta}_{0}/\tilde{\alpha}\right\vert $.

We have performed the linear stability analysis of these solutions
by analysing the linearized evolution of perturbations of the form
$\delta ve^{\lambda t}e^{ikx}$ with $\delta v$ small. The results
can be summarized as follows: (i), the trivial solution, solution $\tilde{\theta}=0$,
and solution $\delta=\pi/2$ are all neutrally stable with two
pairs of complex conjugated purely imaginary eigenvalues (different
for each solution); and (ii), solution $\delta=-\pi/2$ is unstable
versus perturbations with wavenumber $0<k^{2}<\tilde{\theta}_{0}$
when $I<\tilde{\theta}_{0}/\tilde{\alpha}$, and unstable versus all
perturbations (more unstable the larger $k^{2}$ is), when $I>\tilde{\theta}_{0}/\tilde{\alpha}$.
All details are given in Appendix B.

Hence, there is a clear distinction between solution $\delta=\pi/2$,
which is neutrally stable, and solution $\delta=-\pi/2$, which is
unstable versus perturbations with non-null wave-number. This is
reminiscent of the modulational instability occurring in optical fibres
\cite{Boyd}.




\subsection{Solitons as stationary solution of the continuum equation}\label{sec:Continuum}

We now aim to look for localized stationary solutions of the system of differential equations defined in Eqs.~\eqref{eqn:estat2}-\eqref{eqn:estat3}.
Let us first consider the case in which $\delta$ is close to $-\pi/2$ where the modulational instability is expected.
Consider the case of a bright localized structure, such as a bright soliton. Notice first that this type of structure tends assymptotically towards the trivial solution far from its center. In particular, far from the structure center it is verifyed that $\partial_x I \to 0$ and $\partial_x\delta \to 0$, so that we can assume that it reaches a nearly homogeneous solution with very small, but non null, intensity, i.e., solution (iii) above. We conclude that far from the structure $\delta = -\pi/2$ and $I\to 0$, which allows us to conclude from Eq.~\eqref{eqn:estat2} that 
\begin{equation}\label{eq:sigma_asympt}
	\partial_t \sigma = - 2 \tilde \theta_0~,
\end{equation}
which will be valid in any region of space since $\sigma$ is homogeneous ($\partial_x \sigma = 0$). Finally, let us assume that the stationary solutions presents small variations of the phase difference around $-\pi/2$, i.e., $\delta \approx -\pi/2 + \Delta$, where $\Delta$ is a small perturbation.
We can rewrite Eqs.~\eqref{eqn:estat2}-\eqref{eqn:estat3} for the new variable $\Delta$, and taking into account Eq.~\eqref{eq:sigma_asympt} we get
\begin{align}
	\partial_x \Delta & = 2 \tilde \theta_0 (1-\cos \Delta) + 2 \tilde\alpha I \cos^2\Delta~,\\
	\partial_x I & = - 2 I \tilde\theta_0 \sin \Delta + 2 I^2 \tilde \alpha \cos \Delta \sin \Delta ~.
\end{align}
Approximating the trigonometric functions up to first order in $\Delta$ these equations reduce to 
\begin{align}
\partial_{x}\Delta & =2\tilde{\alpha}I~,\\
\partial_{x}I & =-2(\tilde{\theta}_{0}-\tilde{\alpha}I)\Delta I\approx-2I\tilde{\theta}_{0}\Delta ~,
\end{align}
where in the last equation we considered that the term $\tilde \alpha I^{2}\Delta$ is negligible.
The solution of this system of equations is now exact and gives
the following solution, 
\begin{equation}\label{eq:StationaryContinuum}
\begin{split}\Delta(x)=\frac{\tilde{\alpha}}{2}\tanh\left(\frac{\tilde{\alpha}\tilde{\theta}_{0}}{2}x\right),\\
I(x)=\frac{\tilde{\alpha}\tilde{\theta}_{0}}{8}\sech^{2}\left(\frac{\tilde{\alpha}\tilde{\theta}_{0}}{2}x\right)~,
\end{split}
\end{equation}
where we imposed the normalization condition to obtain the constants of integration. This solution represents the usual shape of bright solitons. In the following section we numerically investigate if the predictions made for the continuous limit still hold for the discrete model.

\begin{figure*}
	\centering 
	\includegraphics[width=0.32\linewidth]{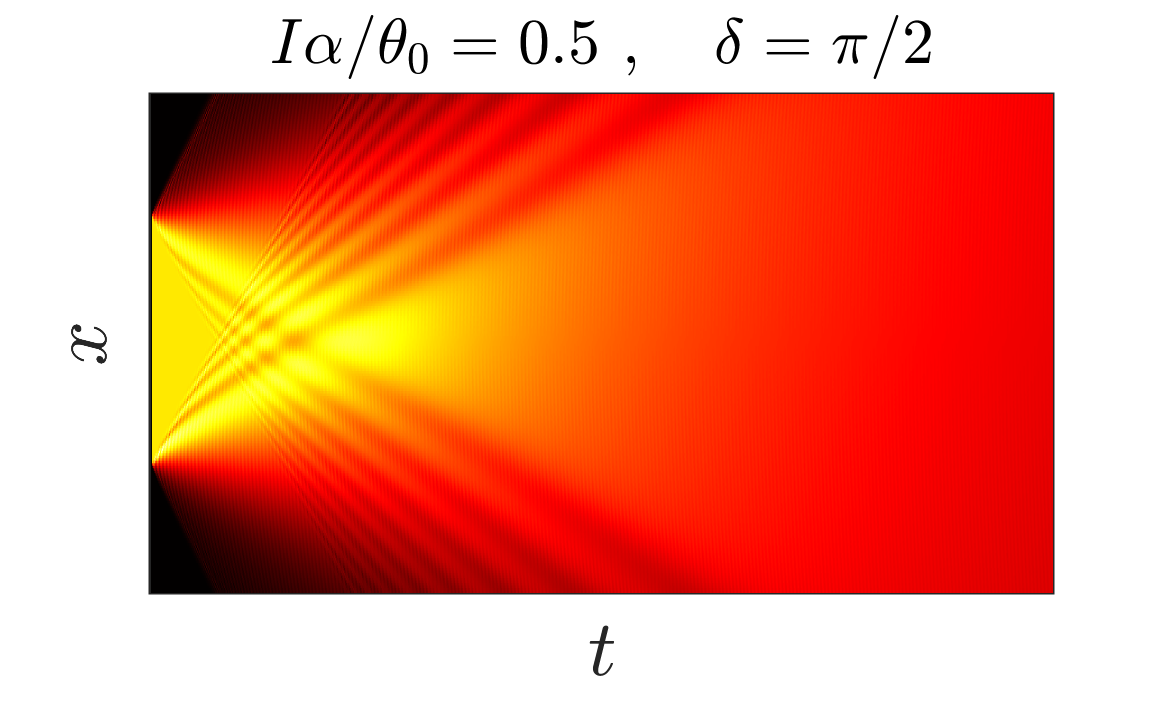}
	\includegraphics[width=0.32\linewidth]{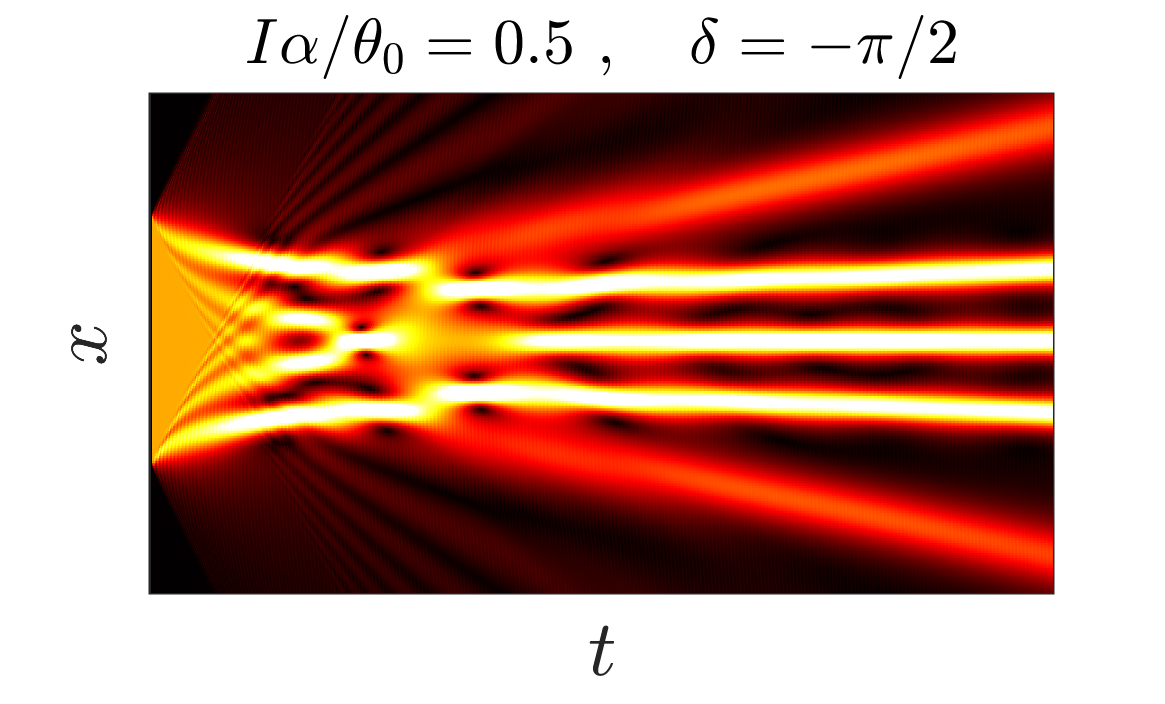}
	\includegraphics[width=0.32\linewidth]{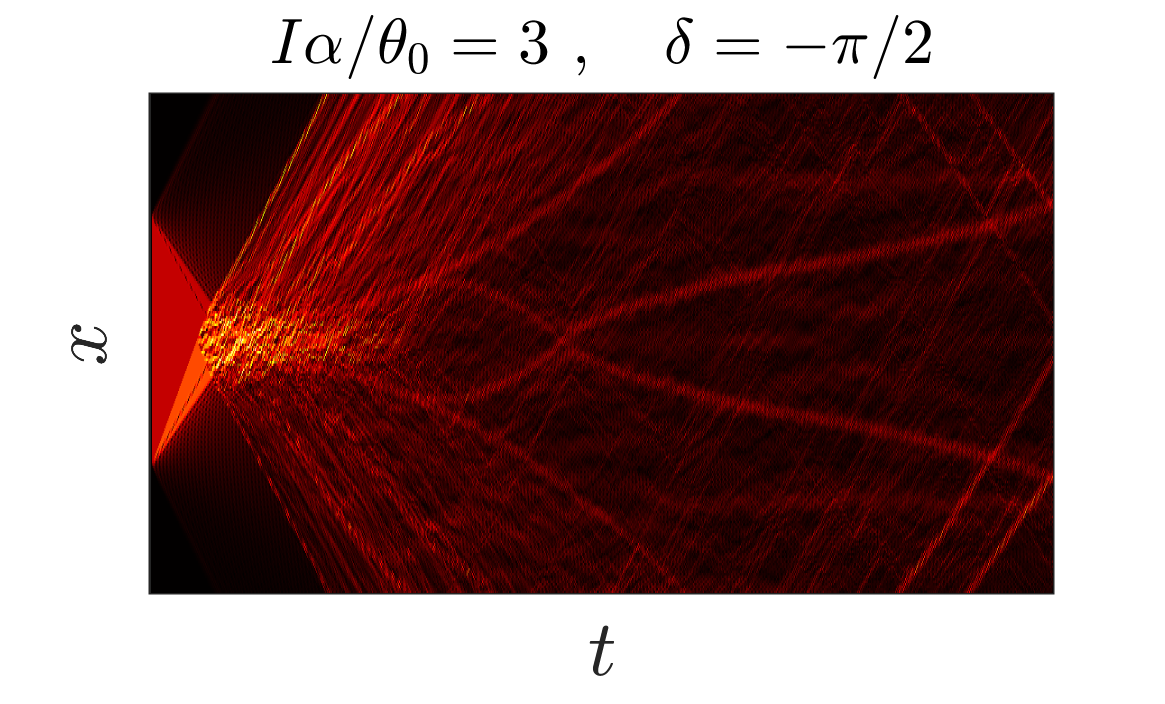}
	\caption{Evolution of an extended initial walker, where the spin states in locations $x=[-50,50]$ all have the same initial coin state. The coin angle is $\theta_{0}=\pi/3$, and the non-linearity parameter is $\alpha=2\pi$ in all panels. 
	In the first panel, the initial coin state is $\ket{\psi_0}=(1,-i)^T/\sqrt{2}$ so that $\delta=\pi/2$, while in the last two panels the initial coin state is $\ket{\psi_0}=(1,i)^T/\sqrt{2}$ with a corresponding $\delta=-\pi/2$. The initial intensity of each initial condition is given in the title of each panel. The intensity or probability density of the walker, defined as $P_{t,x} = |u_{t,x}|^2 + |d_{t,x}|^2$, is given by a heatmap, where black indicates a low probability density and brighter/hotter colours indicate a higher probability density.}
	\label{fig:stab} 
\end{figure*}

As for the case $\delta=\pi/2$, even if there is not a modulational instability, we can still expect the formation of dark solitons. A dark soliton is nothing but a domain wall connecting two domains in which there is a $\pi$ phase difference in the field components, a sign change, which manifests in the intensity as a dark line at the center of the domain wall separating two domains with homogeneous intensity. We can expect the formation of such structures because Eqs.~\eqref{eqn:estat1}-\eqref{eqn:estat3} only depend on the intensity $I$, which means that the field amplitudes can take any of the two values $\pm\sqrt{I}$, thus allowing for the formation of domain walls. We numerically show below that this is actually the case. 

\section{\label{sec:Soliton-formation}Numerical}


The continuum limit of the NLQW map proposed in Eq.~\eqref{eq:MapQWNL} and its stability of analysis of homogeneous stationary solutions predicts the formation of bright and dark solitons. In this section we numerically investigate if these predictions are obeyed by the discrete NLQW and compare the structure of bright solitons with the analytical prediction.

We first explore in Fig.~\ref{fig:stab} the evolution of an extended initial condition where the spin of the walker is uniformly distributed, and consider three distinct types of stable or unstable regimes. 
In the left panel of Fig.~\ref{fig:stab}, the walker is uniformly distributed with components phase difference $\delta=\pi/2$, which according to the stability analysis of Sect. \ref{sec:stability} is a neutrally stable configuration, i.e.,  perturbations are not enhanced nor diminished. It can be seen that the probability density of the walker with this phase difference is mostly uniform after some initial interactions.
When $\delta=-\pi/2$ the stability was dependent on the value of the intensity $I$. On the one hand, when $I\alpha/\theta_0 < 1$ only perturbations with small wave number ($0<k<\theta_0$) are enhanced. In the central panel of Fig.~\ref{fig:stab} we can see the appearance of soliton-like structures that have an extended (low $k$) stable probability distribution. On the other hand, when $I\alpha/\theta_0 > 1$ perturbations of any wave number are enhanced. In the right panel of Fig.~\ref{fig:stab} it can be seen that, for a higher value of the intensity, only very narrow structures, with high wavenumber $k$, are formed.

\begin{figure}
	\centering 
	\includegraphics[width=\linewidth]{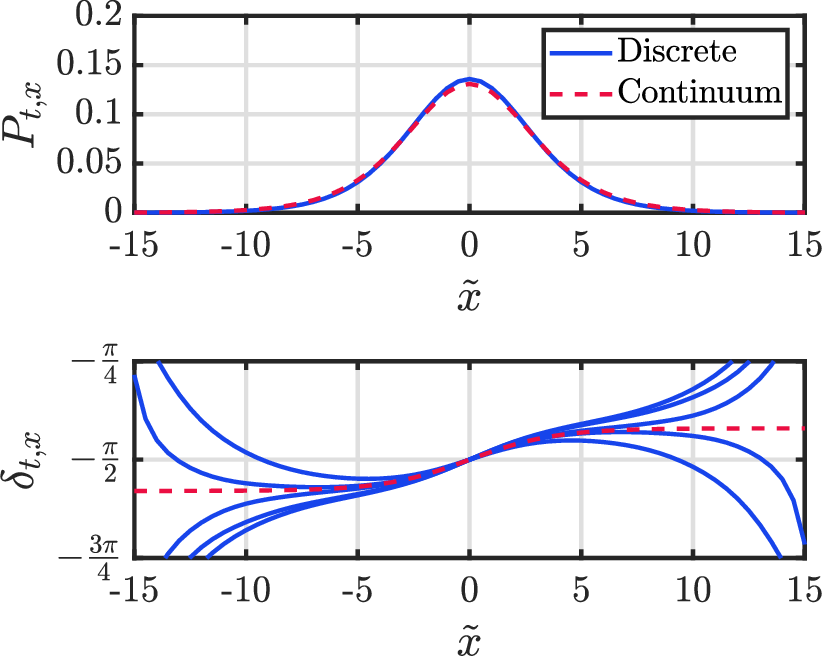}
	\caption{(Top) Probability distribution of the walker. (Bottom) Phase difference between walker components at different consecutive time steps. The red dashed line represents the analytical solution obtained in Eq.~\eqref{eq:StationaryContinuum} for both quantities. The initial condition is \eqref{eq:InitialStatic} evaluated after $t=500$ steps and 4 subsequent steps.
	The parameters of the quantum coin are $\tilde\alpha=1$ and $\tilde\theta_{0}=\pi/3$ and the initial width of the walker is given by $\beta=\tilde\alpha\tilde\theta_0/2$, for a spacing $\epsilon=0.5$. The spatial coordinate has also been scaled as $\tilde x = \epsilon x$. }
	\label{fig:stationary_sol} 
\end{figure}

\subsection{Bright solitons}\label{sec:NumSoliton}

When the soliton-like structures of the central panel of Fig.~\ref{fig:stab} are formed, we obtained that the probability distribution of the walker components are well described by the typical $\sech^2(x)$ function, which was also predicted for stationary solutions of the continuum limit. 
If we consider this distribution as an initial condition with relative phase between walker components $\delta=-\pi/2$ 
\begin{equation}
\Braket{x|\psi_{0,x}^{\mathrm{soliton}}}=N_\beta\begin{pmatrix}\sech(\beta x)\\
i\sech(\beta x)
\end{pmatrix}~,\label{eq:InitialStatic}
\end{equation}
where $N_\beta$ is a normalization constant that depends on $\beta$, we observed that the associated probability distribution remains stationary at different times. 
In Fig.~\ref{fig:stationary_sol} we show the probability distribution of the walker $P_{t,x} = |u_{t,x}|^2 + |d_{t,x}|^2$, and the difference of the walker phases $\delta_{t,x}$ after $t=500$ and 4 subsequent steps. We also plot the stationary solution obtained in the continuum limit Eq.~\eqref{eq:StationaryContinuum}. The probability distribution is stationary and nicely fits the analytical solution. The phase differences have oscillating values around the boundary of the soliton, but the behaviour around the centre of the soliton is well described by the approximate analytical solution of the continuum model.

We also observed that the phase sum is constant along the $x$ direction, while it has a linear dependence in time. This dependence is observed to be 
\begin{equation}
	\sigma_{t,x}^{\mathrm{soliton}} = \sigma_0 - 2 \theta_0 t~,
\end{equation}
where $\sigma_0$ is the initial value of the phases sum, and we notice that this expression is valid in the regions inside the soliton. This observation is in agreement with the results obtained in Sect.~\ref{sec:Continuum}.

\begin{figure}
	\centering 
	\includegraphics[width=1\linewidth]{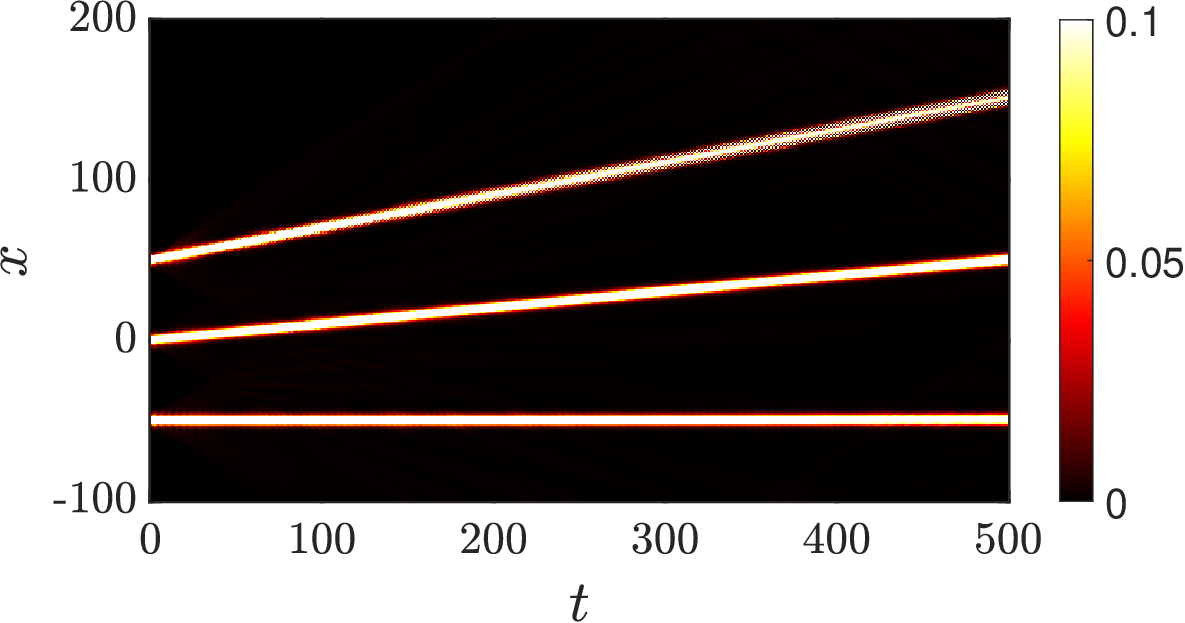}
	\caption{Density plot of the evolved probability distribution of three solitons initially localized at $x=50$, $x=0$ and $x=-50$ with different relative phase distributions: $\nu=2/3$, $\nu=1/2$ and $\nu=0$, respectively. They all have the shame initial width with $\beta=1/2$. The angle is $\theta_{0}=\pi/4$ and the non-linearity parameter is $\alpha=\pi$.}
	\label{fig:soliton-propagation} 
\end{figure}
\begin{figure}
	\centering \includegraphics[width=1\linewidth]{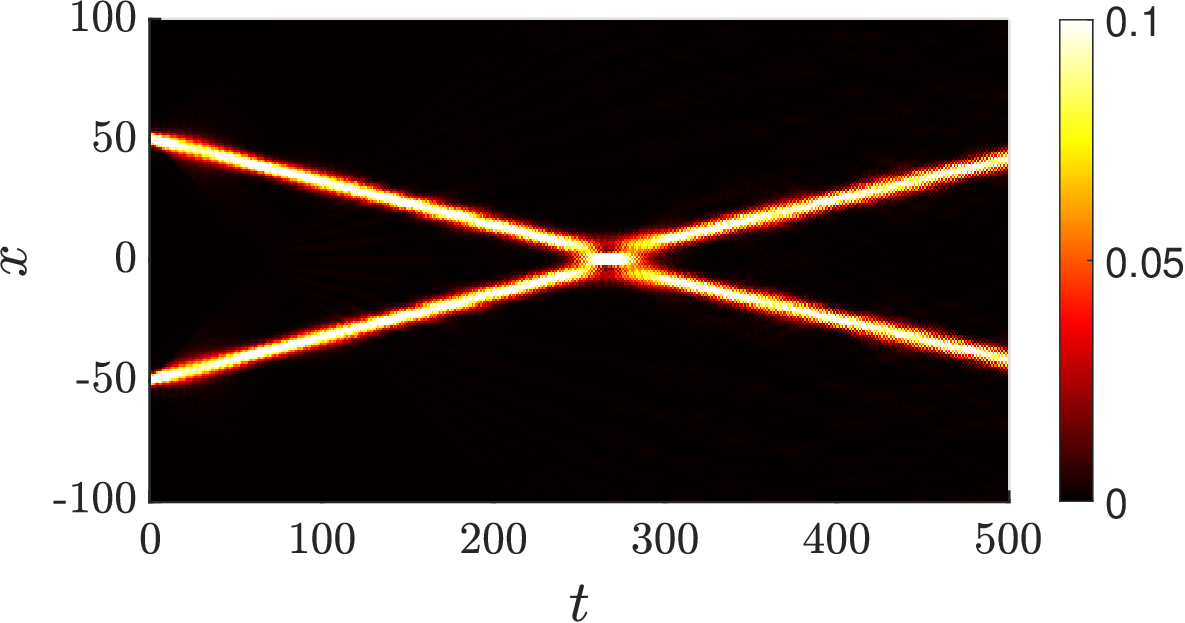}
	\caption{Density plot of the evolved probability distribution of two solitons initially localized at $x=50$ and $x=-50$ that propagate in opposite directions with $\nu=-2/3$ and $\nu=2/3$, respectively. They all have the shame initial width $\beta=1/2$. The angle is $\theta_{0}=\pi/4$ and the non-linearity parameter is $\alpha=\pi$.}
	\label{fig:soliton-colision} 
\end{figure}

When considering an initial condition of the form 
\begin{equation}
\Braket{x|\psi_{0,x}}=N_\beta\begin{pmatrix}\sech(\beta x)\\
ie^{i\nu\tanh(\beta x)}\sech(\beta x)
\end{pmatrix}~,\label{eq:InitialDesp}
\end{equation}
we did not observe a stationary soliton, but a soliton that propagates at a constant velocity.
We observed that if $\nu$ is positive, the initial soliton-like structure propagates to the right (positive $x$) and, if it is negative, it would propagate to the left, i.e., $\nu$ plays the role of velocity on this initial condition. In Fig.~\ref{fig:soliton-propagation} we present the evolution of three initial solitons propagating with different values of $\nu$: two that propagate with different velocities, and another one with $\nu=0$ that remains stationary. The probability distribution and relative phases are the same as in the static soliton, but with the centre displaced at a constant velocity. This initial condition produces a kick, after which he soliton propagates at a constant velocity. 

Another feature that is characteristic of solitons is that the interaction between them leave the shape of their wave packets unaltered. This effect is also showcased by the solitons generated in this QW. In Fig.~\ref{fig:soliton-colision} we show the collision of two solitons propagating in opposite directions, and it can be observed that they cross each other without any significant modification after the crossing.

\begin{figure}
	\centering
	\includegraphics[width=\linewidth]{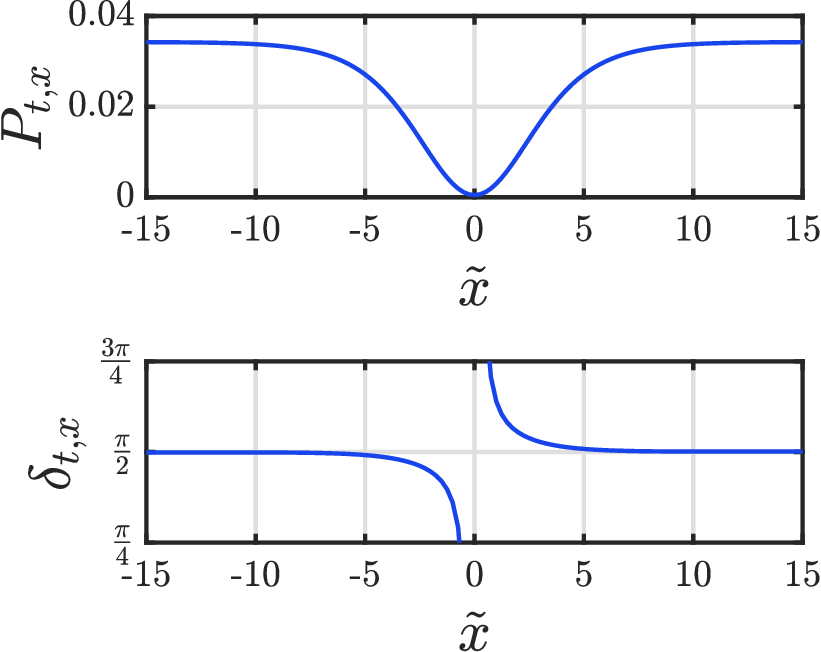}
	\caption{Probability distribution of the walker (upper panel), and phase difference between components (lower panel) with initial condition \eqref{eqn:dark_soliton} evaluated after $t=500$ steps. The parameters of the quantum coin are $\tilde\alpha=1$ and $\tilde\theta_{0}=\pi/3$, the initial width of the walker is given by $\beta=\tilde\alpha\tilde\theta_0/2$, and the intensity $I=\beta$, for a spacing $\epsilon=0.5$. The spatial coordinate has also been scaled as $\tilde x = \epsilon x$. }
	\label{fig:dark_soliton} 
\end{figure}

\begin{figure}
	\centering
	\includegraphics[width=\linewidth]{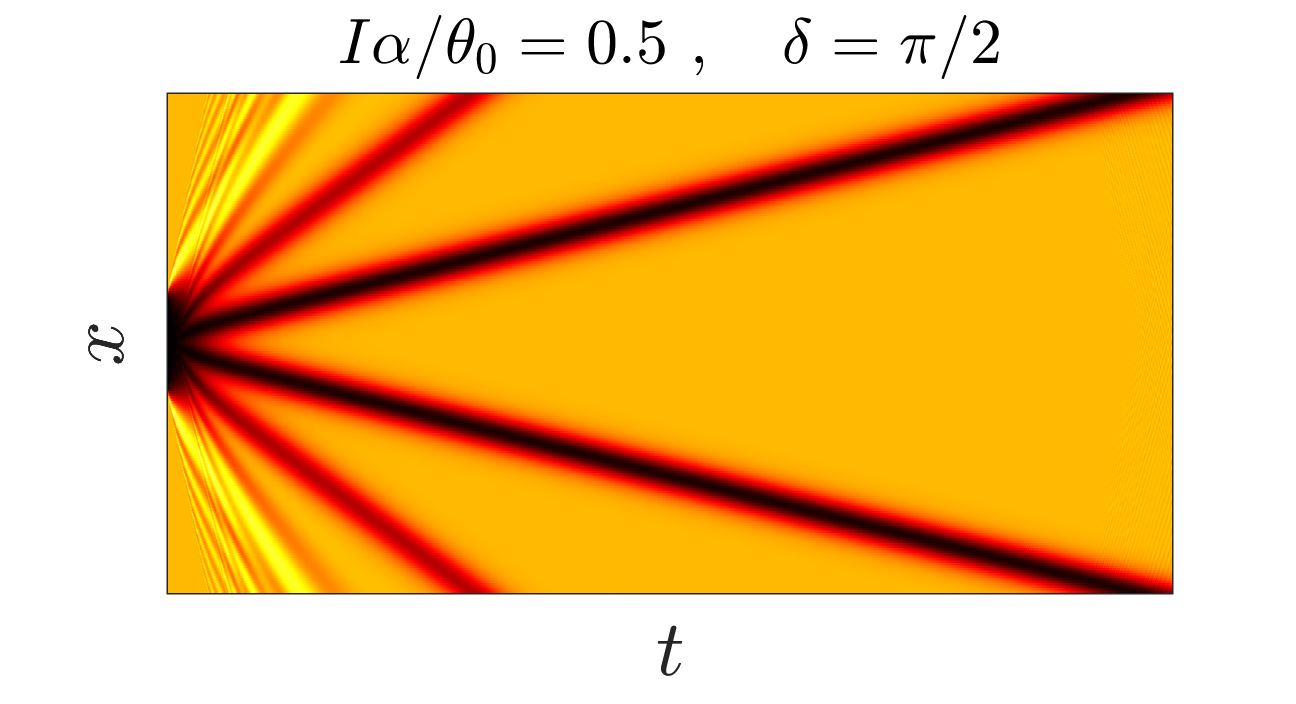}
	\caption{Evolution of the probability distribution of the walker. The probability of finding the walker in locations $x=[-50,50]$ is initially null, while it is constant everywhere else. The initial spin state, where the intensity is constant, is $\ket{\psi_0}=(1,-i)^T/\sqrt{2}$ so that $\delta=\pi/2$, the coin angle is $\theta_{0}=\pi/3$ and the non-linearity parameter is $\alpha=2\pi$. }
	\label{fig:dark_soliton_formation} 
\end{figure}

\begin{figure*}
	\centering 
	\includegraphics[width=0.32\linewidth]{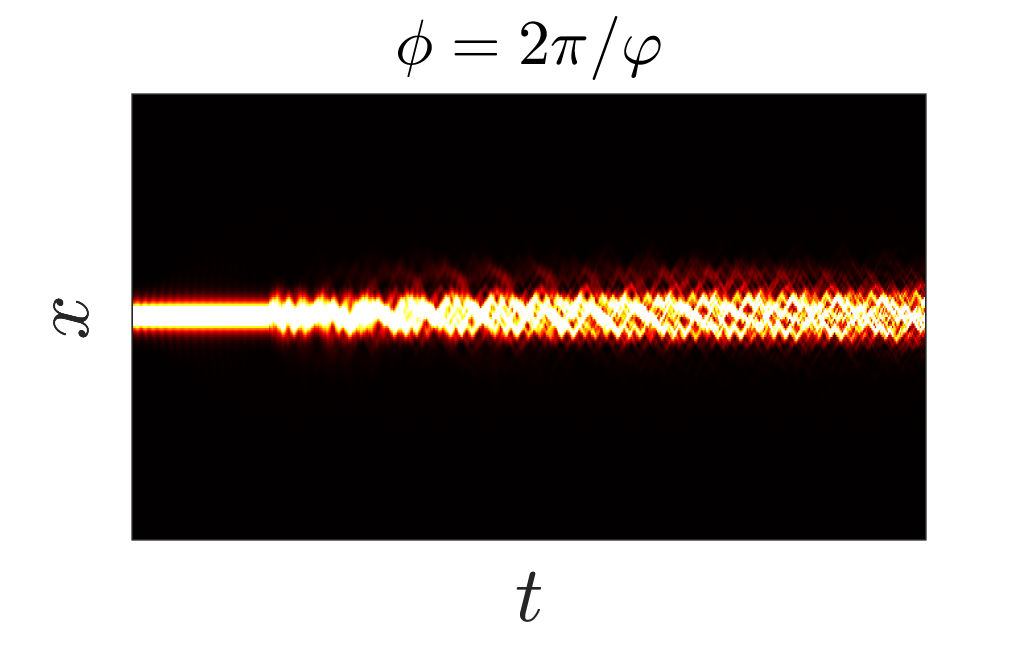}
	\includegraphics[width=0.32\linewidth]{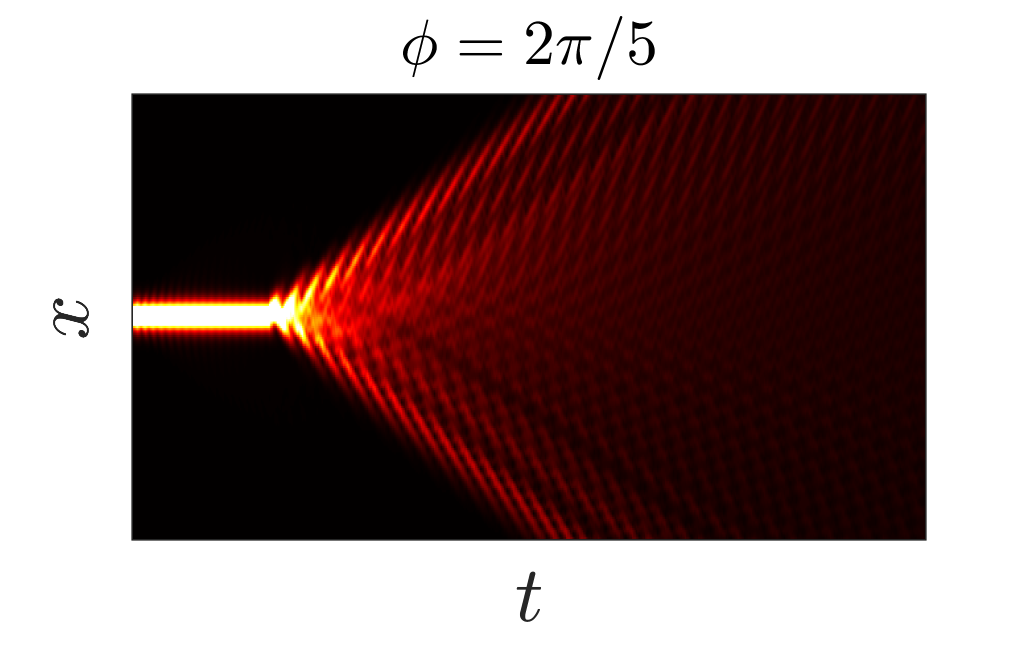}
	\includegraphics[width=0.32\linewidth]{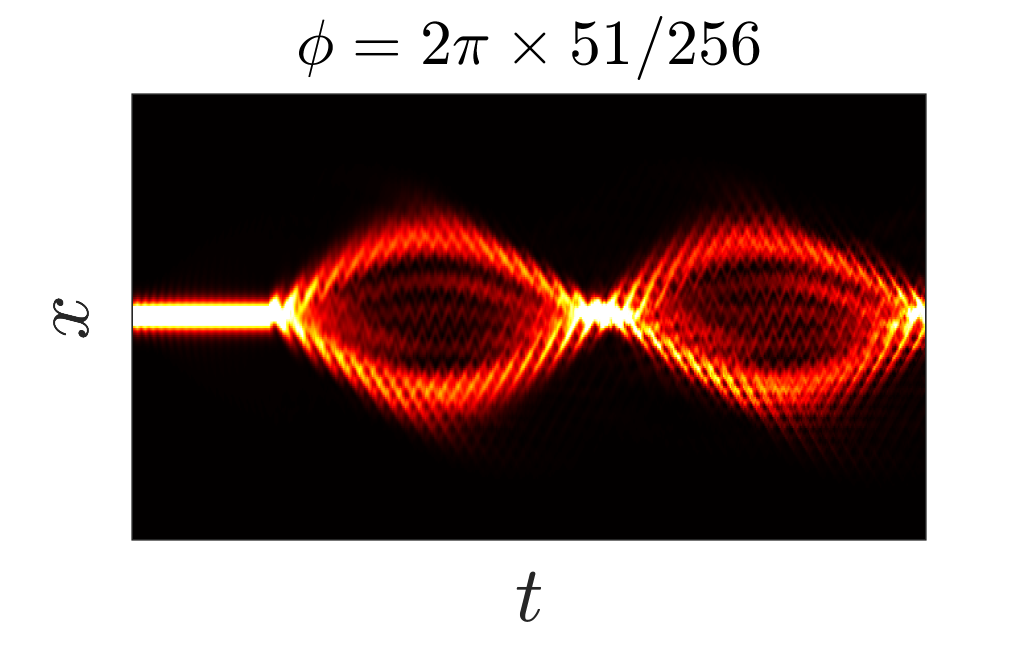}
	\caption{Evolution of the initial soliton considered in Fig.~\ref{fig:stationary_sol} which is subject to a constant electric field after $t=100$.
The probability density of the electric field is given in the title of each panel. The intensity of the walker is given by a heatmap, black indicates low probability density and brighter/hotter colours indicate higher probability density.}
	\label{fig:soliton-electric-prop} 
\end{figure*}

\subsection{Dark solitons}

We saw in the continuum limit that for $\delta = \pi/2$ there is not any modulational instability, as homogeneous solutions are marginally stable. Hence, the formation of bright solitons is not expected to occur in this case. However, as the equations allow for homogeneous states of intensity $I$ with amplitudes $\pm\sqrt{I}$, one can expect the formation of dark solitons, in the regions that connect these two possible amplitudes or solutions.
In Fig.~\ref{fig:dark_soliton} we represent the stationary probability and phase difference for an initial condition
\begin{equation}\label{eqn:dark_soliton}
	\Braket{x|\psi_{t}^{\mathrm{dark}}}=\sqrt{I}\tanh(\beta x)\begin{pmatrix}1\\-i\end{pmatrix}~,
\end{equation}
with a smooth transition between the two regions with opposite amplitude sign. It can be observed that the left and right regions remain constant and keep the initial phase difference of $\delta =\pi/2$. The valley of the central part represents the dark soliton, and right in the centre, where the probability distribution is null, the phase difference is not well-defined. 

In Fig.~\ref{fig:dark_soliton_formation} the formation of many propagating dark solitons, from two homogeneous regions that are initially spatially separated by a dark region, is observed. After some initial interaction around the boundary regions of the initial walker, some domains of constant intensity are formed. These domains are delimited by regions of near null probability density, which we already saw are stable; these are the dark solitions that have the same characteristics as the ones observed in Fig.~\ref{fig:dark_soliton}.

\subsection{Solitons in electric fields}

We now explore whether these structures are robust against the presence of electric fields \cite{DTQWElectric,PhysRevA.73.062304}. To include the effect of an electric field we modify the step evolution defined in Eq.~\eqref{eq:EvolutionUnitary} by 
\begin{equation}
	\ket{\psi_{t+1}}=e^{i\Phi X}SC\ket{\psi_{t}}~,\label{eq:EvolutionUnitaryElectric}
\end{equation}
where $X$ is the position operator, and $\Phi$ plays the role of an electric field intensity. In the limit where the non-linearity parameter $\alpha$ is null, this unitary evolution corresponds to a Dirac equation with constant electric field in the continuum limit. 
For the linear DQW it was pointed out in \cite{DTQWElectric} that if $\Phi$ is an irrational multiple of $2 \pi$, the walker exhibits localization. 
If $\Phi$ is a rational fraction of $2\pi$, i.e. $\Phi=2\pi n/m$, the walker can exhibit oscillations around the initial position but it will eventually become ballistic after a time that will depend on $m$. 
In Fig.~\ref{fig:soliton-electric-prop} we explore the dynamics of a soliton of the NLQW under the effect of the electric field in three regimes: $\Phi=2\pi/\varphi$, $\Phi=2\pi/5$ and $\Phi=2\pi\times 51/256$, with $\varphi=(1+\sqrt{5})/2$ the golden ratio. The first case is known to correspond to the most irrational number, whereas the last two cases give a very close value $\Phi$, the only difference being that the denominator is much larger for the last case. 
It can be observed that in the irrational case the walker remains localized, but the smooth structure of the initial soliton is lost. For the second case, the walker undergoes some oscillations but quickly becomes ballistic. In the last case the soliton is seen to split into two components that undergo oscillations and present some interference patterns. 

We have observed (not shown) that the effect of the electric field dominates over the nonlinear rotation angle. The dynamics of the soliton is very similar to the dynamics of a linear walker with an extended probability distribution subject to an electric field. The effect of the nonlinear angle is only apparent at longer time scales where interferences become dominant. We have also observed (not shown) that the same phenomenology is displayed by dark solitons in the presence of electric fields.


\subsection{No solitons in 2D}

This NLQW can be extended to a two-dimensional spatial Hilbert space $\mathcal{H}_{x}\otimes\mathcal{H}_{y}$ with basis $\{\ket x\otimes\ket y\}$. We will make use of the split-step evolution for this QW \cite{PhysRevA.84.042337,PhysRevA.87.022336}, so that the same coin space and coin operators can be used with spinor components $u_{t,x,y}$, $d_{t,x,y}$. The time step is therefore defined as 
\begin{equation}
\ket{\psi_{t+1}}=S_{y}CS_{x}C\ket{\psi_{t}}~,\label{eq:Evolution2D}
\end{equation}
where $S_{i}=e^{-i\hat{p}_{i}\sigma_{z}}$ is the conditional shift operator in the direction $i=\{x,y\}$, and $C$ is the same coin operators as before, with the rotation angle similarly defined as 
\begin{equation}
\theta_{t,x,y}=\theta_{0}+\alpha|u_{t,x,y}||d_{t,x,y}|\sin(\varphi_{t,x,y}^{u}-\varphi_{t,x,y}^{d})~,\label{eq:NLangle2D}
\end{equation}
where there is a dependence on the values of the walker in both dimensions.

It was discussed in \cite{SearchQWNL} that non-linear QWs, that introduce the nonlinearity in form of phases on the walker components, can be exploited to perform efficient search tasks on the two-dimensional grid. In line with those findings, we observed that the NLQW that introduces nonlinearities in the coin rotation operator angle produces ballistic dispersion, indicating that no soliton-like structures are formed in the two-dimensional case. 

\section{\label{sec:Conclusions}Conclusions}

In this work, we have proposed and analyzed a nonlinear QW model which can be experimentally implemented using the components of the electric field on an optical nonlinear Kerr medium. Differently to the Non Linear Optical Galton Board model proposed in \cite{Navarrete07}, where nonlinearities manifest as a set of different phases of the coin operator (or, equivalently, of the displacement operator), here they give rise to a rotation in the coin operator, with a single angle which depends (in a nonlinear fashion) on the state of the walker. This simple dependence makes it easy to consider the space-time continuum limit of the evolution equation, which takes the form of a nonlinear Dirac equation. The analysis of this continuum limit allows us, under some approximations, to gain some insight into the nature of the soliton structure, which is illustrated by our numerical calculations. 

These solitons are stable structures whose trajectories can be modulated by choosing the appropriate initial condition.
From the continuum limit stability analysis, we were able to predict the existence of both bright and  dark solitons, which were numerically characterized. 
We have also studied the stability of solitons when they are subject to an additional phase that simulates an external electric field, for different rational and irrational values of the field strength.
Finally, we also explored a 2D version of this model, where no evidence of soliton formation was found.


To summarize, nonlinear quantum walks constitute an interesting field with a rich phenomenology that can be used for a better control of its algorithmic and simulation properties. We also remark that the continuum limit of the DQW provided invaluable insight on the properties of the discrete model, which allowed us to predict the existence of both bright and dark solitons.

\begin{acknowledgments}
This work has been founded by the Spanish MCIN/AEI/10.13039/501100011033 grant PID2020-113334GB-I00, SEV-2014-0398 and Generalitat Valenciana grant CIPROM/2022/66, the Ministry of Economic Affairs and Digital Transformation of the Spanish Government through the QUANTUM ENIA project call - QUANTUM SPAIN project, and by the European Union through the Recovery, Transformation and Resilience Plan - NextGenerationEU within the framework of the Digital Spain 2026 Agenda, and by the CSIC Interdisciplinary Thematic Platform (PTI+) on Quantum Technologies (PTI-QTEP+). This project has also received funding from the European Union¿s Horizon 2020 research and innovation program under grant agreement CaLIGOLA MSCA-2021-SE-01-101086123.
\end{acknowledgments}

\appendix

\section{Derivation of the continuum Limit}

\label{app:ContLimit} We can rewrite the walker time step of Eq.~\eqref{eq:EvolutionUnitary}
as 
\begin{equation}
\Psi(t+\epsilon,x)=e^{-\epsilon\partial_{x}\sigma_{z}}e^{-i\sigma_{y}\epsilon\tilde{\theta}(t,x)}\Psi(t,x)
\end{equation}
where we have expressed it in terms of $t$ and $x$ that are both
discretized by the same amount $\epsilon$. When a small $\epsilon$ is taken, the continuum limit of the equation is obtained. We have rescaled the original rotation angle by the same spacing $\tilde{\theta}(t,x)\epsilon=\theta_{t,x}$.
These definitions allow us to write the l.h.s. of the equation as
\begin{equation}
\Psi(t+\epsilon,x)\approx\Psi(t,x)+\epsilon\partial_{t}\Psi(t,x)~,
\end{equation}
while we can approximate the r.h.s. as 
\begin{equation}
\Big(1-\epsilon\sigma_{z}\partial_{x}\Big)\left(1-i\epsilon\tilde{\theta}(t,x)\sigma_{y}\right)\Psi(t,x)~.
\end{equation}
The zero order $O(\epsilon^{0})$ exactly match in both sides, and
the first order terms $O(\epsilon)$ define the following continuous
equation 
\begin{equation}
\partial_{t}\Psi(t,x)=-\sigma_{z}\partial_{x}\Psi(t,x)-i\tilde{\theta}(t,x)\sigma_{y}\Psi(t,x)~,
\end{equation}
which can be rewritten as Eq.~\eqref{eq:DiracNL}.

\section{Stability analysis}

We start with Eqs.~\eqref{ecs1} and substitute in them $f_{i}\left(  x,t\right)  =\bar{f}_i+\delta f_{i}\left(
x,t\right)  $, $f_{i}\in\left\{  |u|,|d|,\delta,\sigma\right\}  $, with $\delta
f_{i}$ small perturbations and $\bar{f}_{i}$ the homogeneous-stationary
solution values. After linearization of Eqs. \eqref{ecs1} around the steady
state (i.e, by neglecting nonlinear terms in the perturbations), the equations
of evolution for the perturbations can be written as%
\begin{align}
\partial_{t}\vec{p} &  =\hat{L}\cdot\vec{p},\label{lin}~,\\
\vec{p} &  =\left(  \delta |u|,\delta |d|,\delta\delta
,\delta\sigma\right)^T~,\\
\hat{L}_{ij} &  =\left[  \frac{\partial\dot{p}_{i}}{\partial p_{j}}\right]_{p_{i}=\bar{p}_{i}}~,
\end{align}
where the matrix elements $\hat{L}_{ij}$ are given by%
\begin{equation}
\begin{split}
	\hat{L}_{11} &  =-\hat{L}_{22}=\left(  \theta_{0}-\bar{\theta}\right) \cos\delta-\partial_{x}~,\\
	\hat{L}_{33} &  =\hat{L}_{44}=\hat{L}_{14}=\hat{L}_{24}=0~,\\
	\hat{L}_{12} &  =\hat{L}_{21}=\left(  \theta_{0}-2\bar{\theta}\right) \cos\bar{\delta}~,\\
	\hat{L}_{13} &  =-\hat{L}_{23}=\sqrt{\bar{I}}\left(  \bar{\theta}\sin \bar{\delta}-\alpha\bar{I}\cos^{2}\bar{\delta}\right)~,\\
	L_{31} &  =-L_{32}=-2\bar{\theta}\sin\delta/\sqrt{\bar{I}}~,\\
	L_{34} &  =-\partial_{x}~,\\
	\hat{L}_{41} &  =\hat{L}_{42}=2\sqrt{\bar{I}}\alpha\sin^{2}\bar{\delta}~,\\
	L_{43} &  =-2\left(  \theta_{0}-2\bar{\theta}\right)  \cos\bar{\delta}-\partial_{x}~,
\end{split}
\end{equation}
with $\bar{\theta}=\theta_{0}+\alpha\bar{I}\sin\bar{\delta}$, and the overbar
indicating homogeneous stationary solutions.

Equation (\ref{lin}) admits solutions of the form
\begin{equation}
\vec{p}_{j}=\vec{p}_{0j}e^{\lambda_{j}t}e^{ikx}~,
\end{equation}
where $\lambda_{j}$ are the eigenvalues of the matrix $\hat{L}$ and $\vec{p}_{j}$ are the eigenvectors. Clearly, whenever $\operatorname{Re}\left(  \lambda_{j}\right)  >0$, for a particular value of $k$, the corresponding steady state is unstable versus perturbations in the form of plane waves with wavenumber $k$.

\begin{figure}
	\centering
	\includegraphics[width=\linewidth]{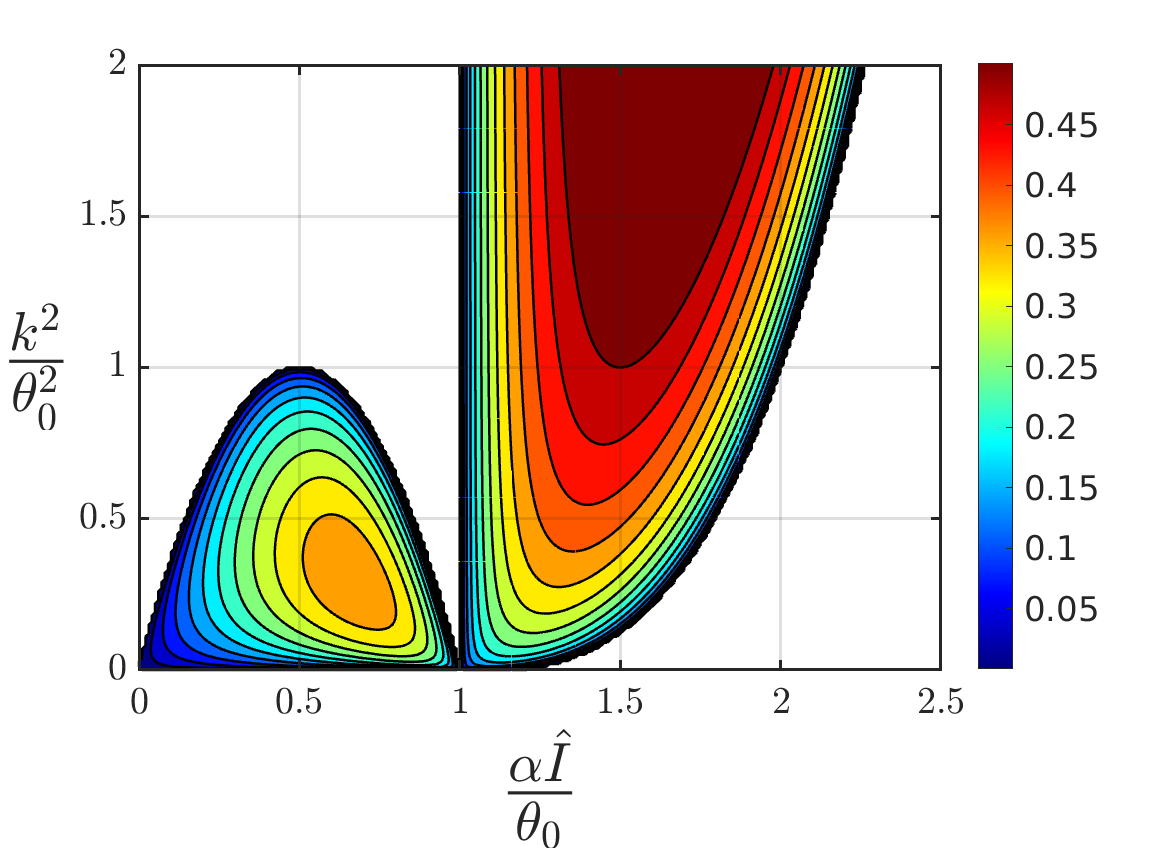}
	\caption{Real part of the eigenvalue that solves the characteristic equation $P_-(\lambda)=0$ and is scaled by $\theta_0$, i.e., we plot $\real(\lambda\theta_0) $. The white regions represent the space where the real part of $\lambda$ is null.}
	\label{fig:stability}
\end{figure}

Next, we detail the stability properties of the different homogeneous steady states. For the trivial solution $\bar{I}=0$, as for solution $\tilde{\theta}=0$, the characteristic polynomial can be written as $\mathcal{P}\left(\lambda\right)  =\left(  \lambda^{2}+k^{2}\right)  ^{2}=0$, hence $\lambda^{2}=\pm ik$ and these solutions are consequently neutrally stable whenever they exist. 

For solutions $\delta=\pm\pi/2$, the characteristic polynomials read, 
\begin{equation}
\begin{split}
	\mathcal{P}_{\pm}\left(  \lambda\right)  &=\lambda^{4}+2\left[  k^{2}+2\left(
\alpha\bar{I}\pm\theta_{0}\right)  ^{2}\right]  \lambda^{2}\\ 
						 &+k^{4}+4k^{2} \alpha\bar{I}\left(  \alpha\bar{I} \pm\theta_{0}\right)  =0~,
\end{split}
\end{equation}
where $\pm$ corresponds to $\pm\pi/2$. It is not difficult to show that $\mathcal{P}_{+}\left(  \lambda\right)  =0$ provides purely imaginary eigenvalues, hence solution $\delta=+\pi/2$ is neutrally stable. On the contrary, $\mathcal{P}_{-}\left(  \lambda\right)  $ provides two couples of eigenvalues, one of which has a positive real part. In Fig.~\ref{fig:stability} we are representing the real part of the eigenvalue, multiplied by $\theta_{0}$, in the plane $\left(\alpha I/\theta_{0},k^{2}/\theta_{0}^{2}\right)$. 
It is clearly seen that: (i) for $\alpha I/\theta_{0}=1$ the eigenvalue is zero for all $k$; (ii) for $\alpha I/\theta_{0}<1$ the eigenvalue is positive for $k^{2}/\theta_{0}^{2}<1$ and null for $k^{2}/\theta_{0}^{2}>1$; and (iii) for $\alpha I/\theta_{0}>1$ the eigenvalue is positive for all $k$. Hence, we conclude that there is a long-wavelength modulational instability whenever $\alpha I/\theta_{0}<1$, and a short-wavelength modulational instability whenever $\alpha I/\theta_{0}>1$.

\bibliography{biblio}
\bibliographystyle{unsrt}

\end{document}